\documentclass[twocolumn,showpacs,preprintnumbers,amsmath,amssymb]{revtex4}
\usepackage{graphicx}
\usepackage{dcolumn}
\usepackage{color}
\usepackage{bm}
\begin{document}
\title {Memory in nanomagnetic systems: Superparamagnetism versus Spinglass behavior}
\author{ Malay Bandyopadhyay$^{1}$ and Sushanta Dattagupta$^{1,2}$}
\affiliation{$^1$S.N. Bose National Centre for Basic Sciences,JD Block, Sector III, Salt Lake, Kolkata 700098, India.\\
$^2$Jawaharlal Neheru Centre for Advanced Scientific Research, Jakkur, Bangalore 560064, India.}
\date{\today}
\begin{abstract}
The slow dynamics and concomitant memory (aging) effects seen in nanomagnetic systems are analyzed on the basis of two separate paradigms : superparamagnets and spinglasses. It is argued that in a large class of aging phenomena it suffices to invoke superparamagnetic relaxation of individual single domain particles but with a distribution of their sizes. Cases in which interactions and randomness are important in view of distinctive experimental signatures, are also discussed.
\end{abstract}
\maketitle
{\section {Introduction}}
 The subjects of both superparamagnetism and spinglasses are quite old and well studied \cite{neel,dorfman,bean,wohlfarth,brown,bouchaud,mezard}. Yet they have been rejuvenated in recent years in the context of fascinating memory and aging properties of nanomagnets. These properties, which are believed to be of great practical usages, have been recently investigated in a large number of experiments on magnetic nanoparticles \cite{sun,chakraverty,kundu,sasaki,zheng,chen,tsoi,raj}. The observed slow dynamical behavior has been variously interpreted, based on the paradigm of either superparamagnet or spinglass, sometimes even obscuring the difference between the two distinct physical phenomena. The purpose of this paper is to reexamine some of the data, others' as well as our own, and critically assess the applicability of the physics of either superparamagnets or spinglasses and occasionally, even a juxtaposition of the two. Our main point is, spinglasses are marked by Complexity, arising out of two separate attributes ----- Frustration and Disorder. While the manifested properties, such as stretched exponential relaxation and concomitant aging effects, can also occur due to `freezing' of superparamagnetism, especially in a polydisperse sample, the physics of spinglasses is naturally much richer than that of superparamagnets. A discernible experimental signature of superparamagnetism versus spinglass behavior seems to be the magnitude of the field-cooled (FC) magnetization memory effect that is significantly larger for the interacting glassy systems than the one in non-interacting superparamagnetic particles \cite{gil}. Therefore, invoking spinglass physics in interpreting data on the slow dynamics of nanomagnets can sometimes be like `killing a fly with a sledge hammer', especially if a simpler interpretation  on the basis of superparamagnetism is available. We explore such situations in this paper.\\
\indent
Superparamagnetism was discussed quite early by Frenkel and Dorfman and later by Kittel, as a property arising out of single-domain behavior when a bulk ferromagnetic or an antiferromagnetic specimen is reduced to a size below about 50 nm
 \cite{dorfman}. For such a small particle-size the domination of surface to bulk interactions yields a mono-domain particle inside which nearly $10^5$ magnetic moments are coherently locked together in a given direction, thus yielding a giant or a supermoment. Clearly, for this to happen, the ambient temperature must be much less than the bulk ordering temperature, so that the integrity of the super moment is maintained. However, as Neel pointed out, in the context of magnetic properties of rocks in Geomagnetism, the direction of the supermoment is not fixed in time \cite{neel}. Indeed, because of thermal fluctuations, this direction can undergo rotational relaxations across an energy-barrier due to the anisotropy of the single-domain particle, governed by the Neel relaxation time:
\vskip-0.3cm
\begin{equation}
\tau = \tau_0 \exp\Big(\frac{KV}{k_BT}\Big).
\end{equation}
In Eq. (1), the preexponential factor $\tau_0$ is of the order of $10^{-9}$ sec, V is the volume of the particle, and K is the anisotropy energy, the origin of which lies in the details of all the microscopic interactions. For our purpose K would be treated as a parameter whose typical value is about $10^{-1}$ Joule/cm$^3$. Therefore, at room temperature, $\tau$ can be as small as $10^{-1}$ sec for a particle of diameter 11.5 nm but can be astoundingly as large as $10^9$ sec for a particle of diameter just about 15 nm. Thus, a slight polydispersity (i.e., a distribution in the volume V), can yield a plethora of time scales, giving rise to interesting slow dynamics. For instance, if $\tau<\tau_E$, where $\tau_E$ is a typical measurement time in a given experiment, the supermoment would have undergone many rotations within the `time-window' of the experiment, thereby averaging out to zero the net magnetic moment. One then has superparamagnetism. On the other hand, if $\tau>\tau_E$, the supermoment hardly has time to rotate within the time-window, thus yielding a `Frozen-moment' behavior. The consequent nonequilibrium features have led to the phrase: ``Magnetic Viscosity" while depicting the time-dependent freezing of moments \cite{bean,wohlfarth,street,sdgbook}. Further, the transition from superparamagnetism to frozen-moment behavior occurs at a temperature, referred to in the literature as the blocking temperature $T_b$, defined by 
\begin{equation}
\tau_E = \tau_0 \exp\Big(\frac{KV}{k_BT_b}\Big).
\end{equation}
When the measurement temperature T is less than $T_b$ the magnetic particles are blocked whereas in the other extreme they display facile response to applied fields. Therefore, we  emphasize that even within a single particle picture, sans any form of inter-particle interactions, such as in a dilute nanomagnetic specimen, one can obtain apparently intriguing effects such as `stretched exponential' relaxation simply because of size distributions. The latter will be shown to be responsible for much of the data on slow relaxations in nanomagnets.\\
\indent
 Turning now to spinglasses, historically the phenomenon was first observed in dilute alloys such as $Au_{1-x}Fe_x$ (or $Cu_{1-x}Mn_x$) in which magnetic impurities Fe (or Mn) in very low concentrations were ``quenched-in" from a solid solution with a host metallic system of Au (or Cu) \cite{mydosh}. The localized spin is coupled with the s-electron of the host metal which itself interacts with the other conduction electrons via what is called the Ruderman-Kittel-Kasuya-Yoshida (RKKY) Hamiltonian, thereby setting up an indirect exchange interaction between the localized moments. Because the coupling constant of the exchange interaction, in view of the RKKY coupling, alternates in sign (between ferro and antiferromagnetic bonds), the system is `frustrated'. Thus the ground state is highly degenerate yielding a zero-temperature entropy. An additional effect is due to disorder. Because the dilute magnetic moments are quenched-in at random sites, the exchange coupling-strengths are randomly distributed. The dual occurrence of frustration and disorder has led to novel concepts in the Statistical Mechanics of spinglasses such as configuration-averaging, replica-techniques (for computing the free energy), broken-ergodicity, etc. \cite{young}. Experimentally, spinglasses are characterized by a `cusp' in the susceptibility and stretched exponential relaxation of time-dependent correlation functions \cite{mydosh}. It is no wonder then that spinglasses also exhibit slow dynamics with associated memory and aging effects, albeit the root causes are much more complex than a system of polydisperse, noninteracting single-domain nanomagnetic particles, discussed earlier. Indeed spinglasses, because of their complexity, have been employed as paradigms for studying real structural glasses, an unresolved problem of modern condensed matter physics \cite{chandan}.\\
\indent
 Given this background on two distinct physical phenomena (and yet manifestly similar properties) of superparamagnets and spinglasses, a natural question to ask is: can there be spinglass-like physics emanating from a collection of single-domain nanomagnetic particles embedded in a non-metallic, non-magnetic host? The answer is clearly an YES when the system is no longer a diluted one such that the supermoments start interacting via dipole-dipole coupling. Because the dipolar interaction (like the RKKY-mediated exchange interaction) is also endowed with competing ferro and antiferromagnetic bonds \cite{luttinger}, as well as randomness due to random locations of the magnetic particles, all the attributes of spinglasses can be simulated in interacting single-domain particles. This will be analyzed below.\\
\indent
With the preceding discussion the plan of the paper is as follows. In Sec. II we discuss the relaxation of non-interacting single-domain particles based on a rudimentary rate theory. The results of this rate theory, coupled with polydispersity of the particles, are applied in Sec. III to a large body of recently published data on the slow dynamics of nanomagnets. Section IV  deals with a different set of experiments that necessarily requires incorporation of interactions between the nanoparticles and hence spinglass-like physics.\\
{\section {Rate theory of relaxation of single-domain particles}}
The Neel formula (Eq. (1)) of relaxation time is the outcome of a generic class of `Escape over Barriers' problems studied by Kramers \cite{kramers}. The supermoment of a single-domain magnetic particle, due to spontaneous thermal fluctuations, is envisaged to undergo rotational Brownian motion across an anisotropy barrier. The latter, in a large class of systems characterized by uniaxial anisotropy, can be described by the energy
\begin{equation}
E(\theta) = KV\sin^2\Theta,
\end{equation}    
where K and V have been introduced earlier, and $\Theta$ is the angle between the anisotropy axis, chosen as Z (determined by the host crystal) and the direction of the supermoment. Thermal fluctuations of the system can be studied in terms of the Fokker-Planck equation for $P(\Theta,t)$, which defines the probability that the supermoment makes an angle $\Theta$ with the anisotropy axis, at a time t \cite{brown,sdgbook}:
\begin{eqnarray}
\frac{\partial}{\partial t}P(\Theta,t)& = &d\frac{1}{\sin\Theta}\frac{\partial}{\partial\Theta}\Big\lbrace\sin\Theta\Big(\frac{1}{k_BT}\frac{\partial E(\Theta)}{\partial\Theta}P(\Theta,t)\nonumber \\
& &+\frac{\partial P(\Theta, t)}{\partial \Theta}\Big\rbrace,\ \ \ \ \ 0\leq\Theta\leq\pi, 
\end{eqnarray}
where d (having the dimension of frequency) is the rotational diffusion constant. Application of Kramers' analysis to Eq. (4) not only yields the Neel relation  (Eq. (1)) but also a set of rate equations in the high barrier / weak noise limit, i.e. $KV>>k_BT$. In this limit the dynamics is basically restricted to $\Theta = 0$ and $\Theta = \pi$ regions and consequently, the Fokker-Planck equation reduces to a set of two-state rate equations:
\begin{equation}
\frac{d}{dt}n_0(t)=-\frac{d}{dt}n_{\pi}(t)=-\lambda_{0 \rightarrow \pi} n_0(t)+\lambda_{\pi \rightarrow 0} n_{\pi}(t),
\end{equation}
where the subscripts on n indicate the two allowed values of $\Theta$ and the rate constants are as follows:
\begin {equation}
\lambda_{0 \rightarrow \pi} = \lambda_{0}\exp[-\frac{V(K+M_s h)}{k_{B}T}],
\end{equation}
$\lambda_{\pi\rightarrow 0}$ being obtained by switching the sign of h. Here $M_s$ is the saturation magnetization per unit volume. Note that Eq. (6) is a generalization of Eq. (1) in order to take cognizance of an external magnetic field h. As discussed in Ref. \cite{chakraverty} the rate equations can be solved analytically for any temperature and field protocol represented by $T(t)$ and $h(t)$, from a given initial condition. For the sake of completeness we rewrite the main result for the time dependent magnetization M(t):
\begin{equation}
M(t) = M(t=0)\exp(-\bar{\lambda}t)\nonumber  + \mu V N \frac{\Delta \lambda}{\bar{\lambda}}[1-\exp(-\bar{\lambda}t)],
\end{equation}
where
\begin{eqnarray}
\bar{\lambda} = \lambda_{0 \rightarrow \pi}+\lambda_{\pi \rightarrow 0};\ \ \ \ \ \Delta{\lambda}=\lambda_{\pi\rightarrow 0}-\lambda_{0 \rightarrow \pi}.
\end{eqnarray}
The observed magnetization of the system is obtained by averaging over a volume distribution 
\begin{equation}
\bar{M}(t) = \int dV P(V) M(t,V).
\end{equation}
The superposition of relaxation rates, caused by the volume distribution P(V), can alter the exponential relaxation indicated in Eq. (7) into a variety of forms, e.g. stretched exponential or logarithmic \cite{street,sdgbook}. Several models of P(V) are extant in the literature, all leading to aging effects. Examples are bimodal distribution \cite{chakraverty}, a flat distribution bounded by two volumes $V_{min}$ and $V_{max}$ \cite{street} or a log-normal distribution (assumed below)
\begin{equation}
P(V) = \frac{\exp\lbrack -\ln(V^2)/(2\gamma^2)\rbrack}{(\gamma V\sqrt{2\pi})},
\end{equation}
$\gamma$ being a fitting parameter.\\
\indent
Until now we have discussed the relaxation effects of isolated (i.e. non-interacting) single-domain nanomagnetic particles. The question we would like to next address is : what happens when these particles are brought closer and the dipolar interaction between their magnetic moments starts becoming non-negligible? Recall that the interaction between two dipole moments $\vec{m}_i$ and $\vec{m}_j$, located at the sites i and j at a distance $|\vec{r}_{ij}|$ apart is given by \cite{abra}
\begin{equation}
{\cal{H}}_{d-d} = \sum_{ij}\gamma_i\gamma_j\frac{\lbrace3\vec{m}_i\cdot\vec{m}_j-(\vec{m}_i\cdot\hat{r}_{ij})(\vec{m}_j\cdot\hat{r}_{ij})\rbrace}{|\vec{r}_{ij}|^3}.
\end{equation}
Here $\gamma_{i}$ and $\gamma_{j}$ are the gyromagnetic ratios of the i$^{th}$ and j$^{th}$ particles respectively, ${\vec r}_{ij}$ is the vector distance between the `sites' at which the two magnetic particles are located, and $\hat{r}_{ij}$ is the corresponding unit vector. It is well known that dipolar couplings, being long-ranged, anisotropic and alternating in the sign of interaction, can indeed lead to frustration and very complex magnetic order of the ground state, depending on the crystal structure \cite{luttinger}. Incorporation of the dipolar interaction into the dynamics is a further complication involving a multi-particle Fokker-Planck equation in which the `drift term', proportional to $E(\Theta)$ in Eq. (4), would have to be replaced by Eq. (10). The underlying theory is quite daunting and is not attempted here, as a simpler treatment is possible for nanoparticles with large anisotropy, as is sketched below.\\
\indent
Recall that the rate equation abstraction of the Fokker-Planck equation is itself a discrete version of a continuous stochastic process, applicable in the high barrier/weak noise limit when the basins of dynamical attractors are restricted to the $\Theta = 0$ and $\Theta = \pi$ regions. In the context of the dipolar coupling, which is after all an anisotropic Heisenberg interaction, this approximation implies that we are in the so-called Ising limit. In this, only the Z-components of the magnetic moments are relevant. The dipolar interaction can now be described by its truncated form \cite{abra}:
\begin{equation}
{\cal H}_{d-d} = {\mu}^2 V^2 \sum_{ij}\gamma_{i}\gamma_{j}\hbar^{2} {\frac{(1-3 {\cos}^2\theta_{ij})}{|\vec {r_{ij}}|^3}} \cos\Theta_{i} \cos\Theta_{j},
\end{equation}
where $\theta_{ij}$ is the angle between $\vec{r}_{ij}$ and the anisotropy axis and $\vec{m}_i$ is replaced by $\mu V\cos\Theta_i$, $\mu$ being the magnetic moment per unit volume and $\Theta_i$ defined after Eq. (3).
The Fokker-Planck dynamics including Eq. (11) is still very formidable. For our purpose we invoke a mean field theory in which the i$^{th}$ nanoparticle say, is envisaged to be embedded in an effective medium that creates a local mean field (MF) at the site i which is proportional to the average magnetization itself. Therefore, ${\cal{H}}_{d-d}$ in Eq. (11) may be replaced by its MF form
\begin{equation}
{{\cal H}^{MF}_{d-d} } =\gamma \hbar{\mu}^2 V^2 \cos\Theta \sum_{j}\gamma_{j}\hbar {\frac{(1-3 {\cos}^2\theta_{ij})}{|\vec {r_{ij}}|^3}} \langle \cos\Theta_{j} \rangle,
\end{equation}
wherein the angular brackets $\langle ... \rangle$ represent thermal averaging. In addition, and in conformity with our stated assumption about the largeness of the anisotropy energy, $\cos\Theta$ can be replaced by the two-state Ising variable $\sigma$ : 
\begin{equation}
{{\cal H}^{MF}_{d-d}} =\gamma {\mu}^2 V^2 \sigma \sum_{j}\gamma_{j}\hbar{\frac{(1-3 {\cos}^2\theta_{ij})}{|\vec {r_{ij}}|^3}} \langle \sigma_{j} \rangle.
\end{equation}
If the nanoparticles are located at random sites of the host matrix, such as grown by the sol-gel technique (for instance $NiFe_{2}O_{4}$ magnetic particles in a $SiO_2$ host \cite{chakraverty}), the interaction in Eq. (13) is random because of random values of $|\vec{r}_{ij}|$ and is also alternating in sign due to different allowed values of $\theta_{ij}$. Within the spirit of the mean field theory the local field H is to be derived self consistently from the following expression:
\begin{equation}
H=\mu \Lambda V \tanh ({\frac{\mu V H}{k_BT}}),
\end{equation}
where $\Lambda$ is a random variable \cite{chakraverty}. Since the local field can point either along $\Theta=0$ or $\Theta=\pi$ direction, within the two-state model, Eq. (14) naturally admits both positive and negative solutions for H.\\
\indent
Summarizing, the effect of interaction within the simplified mean field approximation, enumerated above, is to modify the rate theory in which the rate constant in Eq. (6) is replaced by
\begin {equation}
\lambda_{0 \rightarrow \pi} = \lambda_{0}\exp[-\frac{V(K+M_sh+H)}{k_{B}T}].\\
\end{equation}
Clearly Eq. (15) is an extremely crude attempt to incorporate dipole-dipole interaction into the dynamics of nanomagnetic particles, and is therefore, expected to have limited validity. The actual spinglass dynamics is a much more complex subject that requires application of sophisticated theoretical tools \cite{fischer}. Yet we find that the simple-minded extension of two-state rate theory, as encapsulated in Eq. (15), is adequate to interpret aging data in interacting systems (Sec. IV).\\ 
{\section {Superparamagnetic Slow Dynamics}}
Recently Sun et al have made a series of measurements on a permalloy ($Ni_{81}Fe_{19}$) nanoparticle sample which demonstrate striking memory effects in the dc magnetization \cite{sun}. These involve field-cooled (FC) and zero-field cooled (ZFC) relaxation measurements under the influence of temperature and field changes. We have also observed very similar memory effects in $NiFe_2O_4$ magnetic particles in a $SiO_2$ host \cite{chakraverty}. More recently Sasaki et al \cite{sasaki} and Tsoi et al \cite{tsoi} have reported similar results for the noninteracting (or weakly interacting) superparamagnetic system of $\gamma-Fe_2O_3$ nanoparticles and ferritin (Fe-N) nanoparticles respectively. Further, to understand the mechanisms of the experimental approach of Sun et al, Zheng et al \cite{zheng} replicated the experiments on a dilute magnetic fluid with Co particles and observed similar phenomena. In this section we present a comparison of simulated results with all the above mentioned experimental observations on the basis of our simple two-state noninteracting model plus a log-normal distribution of particle size, described in Sec. II.\\
\vskip -2.0cm
\begin{figure}[h]
{\rotatebox{0}{\resizebox{8cm}{8cm}{\includegraphics{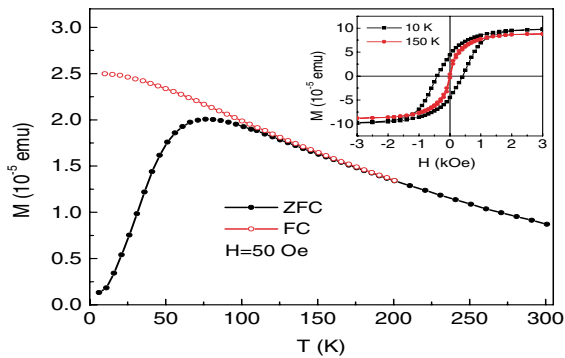}}}}
\vskip-2.0cm
\caption{(color online) Temperature dependence of the dc magnetization for the FC and ZFC processes. Inset shows the M-H curves below and above $T_b$. (Sun et al Phys. Rev. Lett. {\em 91}, 167206 (2003)).}
\end{figure}
We begin our discussion from the most basic and well known protocol, viz. the measurement of the zero field-cooled magnetization (ZFCM) and the field-cooled magnetization (FCM). Figure 1 shows the ZFCM and FCM curves in a 50 Oe field for $Ni_{81}Fe_{19}$. The ZFCM has a peak at $T_{max} = 78K$, which corresponds to the blocking temperature $T_b$. The magnetization of the FC curve continues to increase with decreasing temperature as would be expected for a system in thermal equilibrium. The two curves depart from one another at a temperature higher than $T_{max}$. The inset shows the M-H curve below and above the blocking temperature. Figure 2 and Figure 3 show the simulated FC-ZFC curves and the M-H curve respectively. Our simulations, based on the two-state noninteracting model, match well with the experimental results.\\
\begin{figure}[h]
{\rotatebox{270}{\resizebox{3cm}{6cm}{\includegraphics{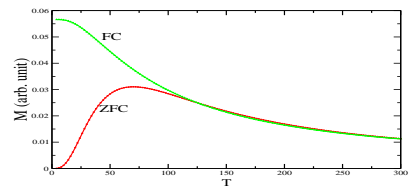}}}}
\caption{(color online) Numerically calculated dc magnetization for the FC and ZFC process.}
\end{figure}
\vskip -0.8cm
\begin{figure}[h]
{\rotatebox{270}{\resizebox{8cm}{8cm}{\includegraphics{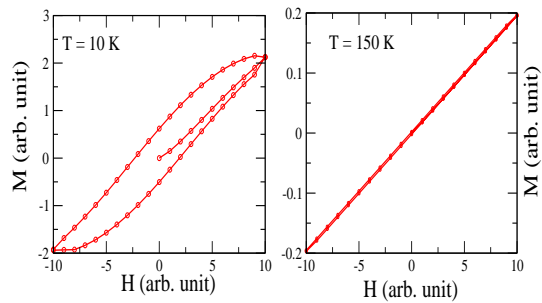}}}}
\vskip-1.9cm
\caption{(color online) Numerically calculated M Versus H curve below and above  $T_b$.}
\end{figure}
The most striking experimental observation of Sun et al is the memory effect in the dc magnetization (Fig. 4) obtained from the following procedure. The sample is cooled in 50 Oe field at a constant cooling rate of 2K/min from 200K ($T_H$) to 10K ($T_{base}$). After reaching $T_{base}$, the sample is heated continuously at the same rate to $T_H$. The obtained M(T) curve is the normal FC curve which is referred to as the reference curve. Then the sample is cooled again at the same rate, but the cooling is arrested three times (at T = 70, 50 and 30K) below $T_b$ with a wait of $t_w = 4h$ at each stop. During $t_w$, the applied field is also turned off to let the magnetization decay. After each stop and wait period, the 50 Oe field is reapplied and cooling is resumed. The cooling procedure produces a step like M(T) curve. After reaching the base temperature, the sample is warmed continuously at the same rate to $T_H$ in the continual presence of the 50 Oe field. Surprisingly, the M(T) curve obtained in this way also shows the step like behavior. Similar memory effects, following the same protocol were seen by us in $NiFe_2O_4$ sample in which the magnetic particles were embedded in a host $SiO_2$ matrix \cite{chakraverty}. The effects can be explained in terms of a bimodal distribution of particle size (i.e. P(V) a sum of two delta functions at volumes $V_1$ and $V_2$) \cite{chakraverty}. Our simulated results based on the two state non-interacting model but accompanied by a log normal distribution, are represented in Fig. 5, which indicate satisfactory agreement with experiments. \\
\vskip -2.0cm
\begin{figure}[h]
{\rotatebox{0}{\resizebox{8cm}{8cm}{\includegraphics{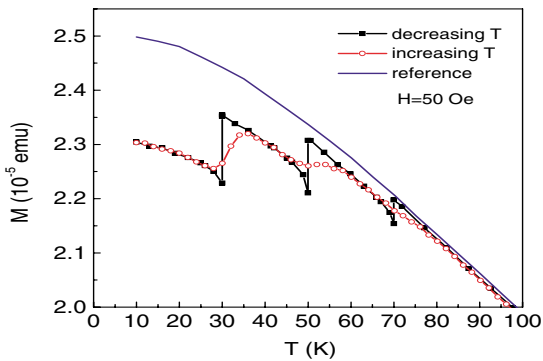}}}}
\vskip-1.0cm
\caption{(color online) ``Memory effect" observed in the dc magnetization measurements in $Ni_{81}Fe_{19}$. (Sun et al Phys. Rev. Lett. {\em 91}, 167206 (2003)).}
\end{figure}
\vskip -0.7cm
\begin{figure}[h]
{\rotatebox{270}{\resizebox{8cm}{8cm}{\includegraphics{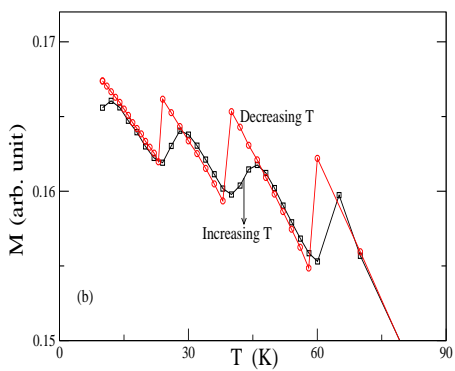}}}}
\vskip-1.0cm
\caption{(color online) Numerically simulated memory effect observed in dc magnetization curves.}
\end{figure}
We further discuss the memory effect observed in  ZFC response measurements (Fig. 6). In the ZFC experiment, the sample is cooled down to $T_0 = 30K$ in zero field. Then a field of 50 Oe is applied and the magnetization is recorded as a function of time. After a time $t_1$, the sample is quenched to a lower temperature, $T_0-\Delta T = 22K$, and the magnetization is recorded for time $t_2$. Finally the temperature is turned back to $T_0$ and the magnetization is recorded for another period $t_3$. The field of 50 Oe is kept on during the entire aging process. When the field is first turned on, a slow logarithmic relaxation takes place following an immediate jump. During the temporary cooling the relaxation is rather weak. When the temperature returns back to $T_0$, viz. 30K, the magnetization comes back to the level it reached before temporary cooling. Moreover it is found that the relaxation curve during $t_3$ is a continuation of the curve during $t_1$.\\
\vskip-2.0cm
\begin{figure}[h]
{\rotatebox{0}{\resizebox{8cm}{8cm}{\includegraphics{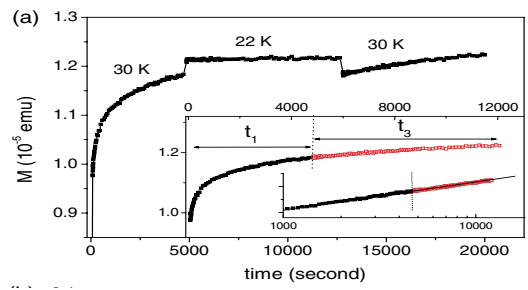}}}}
\vskip-1.5cm
\caption{(color online) Magnetic relaxation measurements in $Ni_{81}Fe_{19}$ with temporary cooling for the ZFC method. Inset shows the same data vs the total time spent at 30 K for both normal and logarithmic time scales.(Sun et al Phys. Rev. Lett. {\em 91}, 167206 (2003)).}
\end{figure}
\vskip-0.5cm
\begin{figure}[h]
{\rotatebox{0}{\resizebox{8cm}{8cm}{\includegraphics{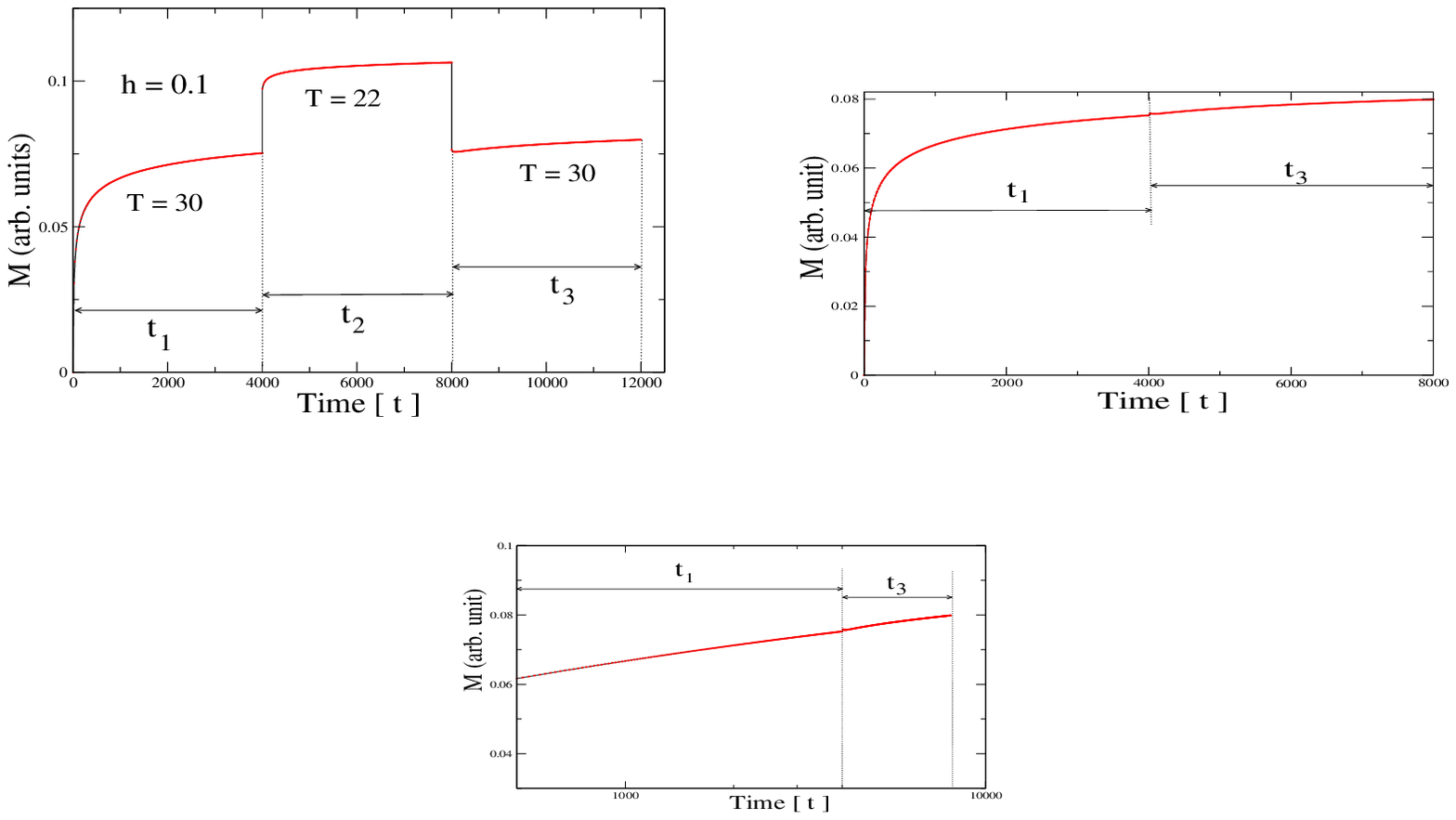}}}}
\caption{(color online)(a) Numerically simulated ZFC relaxation curves with temporary cooling; the same data vs the total time spent at 30 K (b) on a normal scale; (c) on a logarithmic scale.}
\end{figure}
In Fig. 7 we show the numerically calculated results of ZFC relaxation, again on the basis of the two-state non-interacting model, which are qualitatively similar to those in experimental measurements (Fig. 6). When a field of 50 Oe is applied, the magnetization reaches a certain value which is determined by the particles with $T_b\leq 30K$. Then a logarithmic relaxation begins which is due to those particles whose $T_b$ are higher than 30 K \cite{tejada}. The sudden increase in magnetization during $t_2$ is due to the particles with $T_b\leq 22K$ which had flipped during $t_1$ in order to reach their new equilibrium state at $T = 22K$. On the other hand the particles with $T_b>30K$ are not in  equilibrium state and relax extremely slowly at $22K$ to yield an almost constant curve during $t_2$. Finally, when the sample is heated back to $30K$, the particles with $T_b\leq 30K$ and those flipped during $t_1+t_2$, return back to the pre-quenching equilibrium state and therefore the relaxation during $t_3$ is the continuation of the curve in $t_1$.\\
\indent  
In Fig. 8 we show the relaxation measurements of Sun et al in the FC method with temporary cooling, in which the sample is cooled to $T_0=30K$ in a 50 Oe field and the relaxation is measured for a time $t_1$ after the field is cut-off. The sample is quenched to $T=22K$, and the magnetization is recorded for time $t_2$. Finally, the temperature is turned back to $T_0$ and the magnetization is recorded for a time $t_3$.\\
\vskip-2.0cm
\begin{figure}[h]
{\rotatebox{0}{\resizebox{8cm}{8cm}{\includegraphics{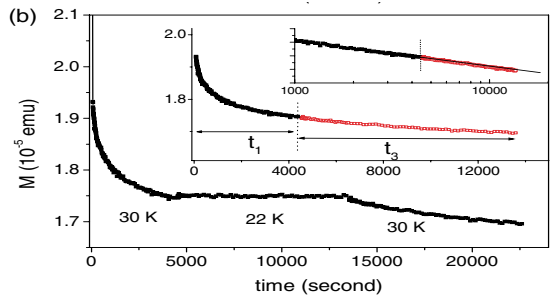}}}}
\vskip-1.5cm
\caption{(color online) Magnetic relaxation measurements in $Ni_{81}Fe_{19}$ with temporary cooling for the FC method. Inset shows the same data vs the total time spent at 30 K for both normal and logarithmic time scales. (Sun et al Phys. Rev. Lett. {\em 91}, 167206 (2003)).}
\end{figure}
\begin{figure}[h]
{\rotatebox{0}{\resizebox{8cm}{8cm}{\includegraphics{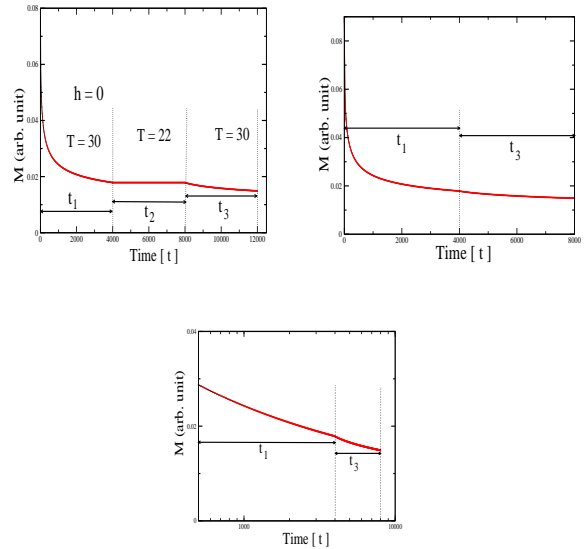}}}}
\caption{(color online) (a) Numerically simulated FC relaxation curves with temporary cooling; the same data vs the total time spent at 30 K (b) on a normal scale; (c) on a logarithmic scale.}
\end{figure}
Fig. 9 shows our numerically simulated results of FC relaxation measurements based on the same two-state individual particle model. After the field is cut-off, the particles with $T_b\leq30K$ show facile response but their contribution to the magnetization is negligibly small, because in their equilibrium state they have almost zero magnetization. So there is a sharp initial drop in M. The subsequent logarithmic relaxation is due to the particles with $T_b>30K$. After $t_1$, the sample is quenched to 22K. Now the remanent magnetization increases slightly, because of the reduced thermal agitation. Since the particles with $T_b>30K$ are not in equilibrium, they relax extremely slow at 22K. Thus we get an almost constant curve during $t_2$. Finally, when the sample is heated to $30K$ the particles with $T_b<30K$ and those flipped during $t_1+t_2$ come back to their previous equilibrium state, and the relaxation in $t_3$ is continuation of the curve in $t_1$.\\
\indent
Figure 10 represents the Sun et al measurements for magnetic relaxation with temporary cooling and field change for the ZFC method. In this, the sample is cooled to $T_0 = 30K$ in zero field. Then a 50 Oe field is applied and the magnetization is measured for a time $t_1$. After $t_1$, the sample is quenched to temperature $T=22K$ in the absence of an external field and the magnetization is recorded for a time $t_2$. Finally the temperature is returned back to $T_0$ and the field is turned on again. The magnetization is measured for a time $t_3$.\\
\begin{figure}[h]
{\rotatebox{0}{\resizebox{8cm}{8cm}{\includegraphics{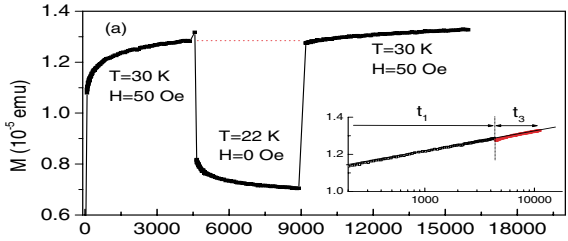}}}}
\vskip-1.5cm
\caption{(color online) Magnetic relaxation measurements in $Ni_{81}Fe_{19}$ with temporary cooling and field change for the ZFC method. Inset shows the same data vs the total time spent at 30 K on a logarithmic time scale. (Sun et al Phys. Rev. Lett. {\em 91}, 167206 (2003))}
\end{figure}
In Fig. 11 we show our corresponding numerically simulated results. When a field of 50 Oe is applied the magnetization immediately reaches a certain value, because the particles with $T_b\leq30K$ equilibrate rapidly. Then a slow logarithmic response begins which is due to the energy distribution of the particles. Now as the field is turned off, we observe a sharp jump in $M(t)$ due to those particles with $T_b\leq22K$ which reach their equilibrium state at $T=22K$ and hence do not contribute to the magnetization. However the particles with $T_b>30K$ are not in equilibrium and relax extremely slow at $T=22K$, so we get a constant curve during $t_2$. Now as the field is turned on again and the temperature of the sample is increased to $T=30K$, the particles with $T_b\leq 30K$ and those flipped during time $t_1+t_2$ come back to the new equilibrium state which is same as that pertaining before quenching. Therefore the relaxation in $t_3$ is the continuation of that during the time $t_1$.\\
\vskip-0.3cm
\begin{figure}[h]
{\rotatebox{0}{\resizebox{8cm}{4cm}{\includegraphics{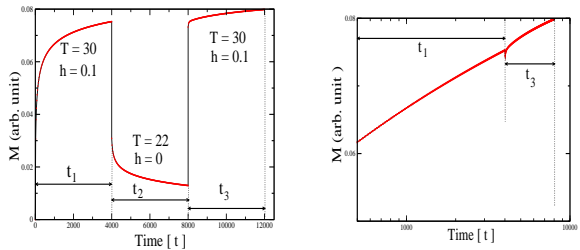}}}}
\caption{(color online) (a) Numerically simulated ZFC relaxation curves with temporary cooling and field change; (b) the same data vs the total time spent at 30 K on a logarithmic scale.}
\end{figure}
\begin{figure}[h]
{\rotatebox{0}{\resizebox{8cm}{8cm}{\includegraphics{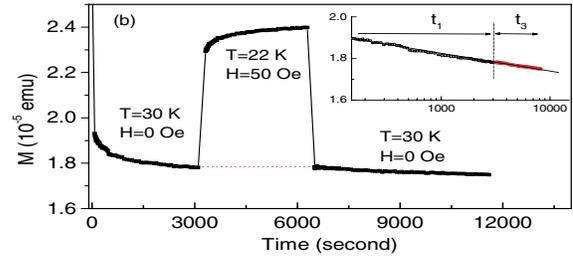}}}}
\vskip-1.5cm
\caption{(color online) Magnetic relaxation measurements in $Ni_{81}Fe_{19}$ with temporary cooling and field change for the FC method. Inset shows the same data vs the total time spent at 30 K on a logarithmic time scale. (Sun et al Phys. Rev. Lett. {\em 91}, 167206 (2003)).}
\end{figure}
Figure 12 shows the experimental results of Sun et al of the magnetic relaxation with temporary cooling and field change in the FC method. In the latter, the sample is cooled to $T_0=30K$ in a 50 Oe field and then the relaxation is measured for a time $t_1$ after the field is cut-off. The field is turned on again and the sample is cooled to $T = 22K$ and the magnetization is recorded for a time $t_2$. Finally the temperature is turned back to $T_0$ and the field is switched off again. The relaxation is now measured for a time $t_3$.\\
\vskip-0.4cm
\begin{figure}[h]
{\rotatebox{0}{\resizebox{10cm}{6cm}{\includegraphics{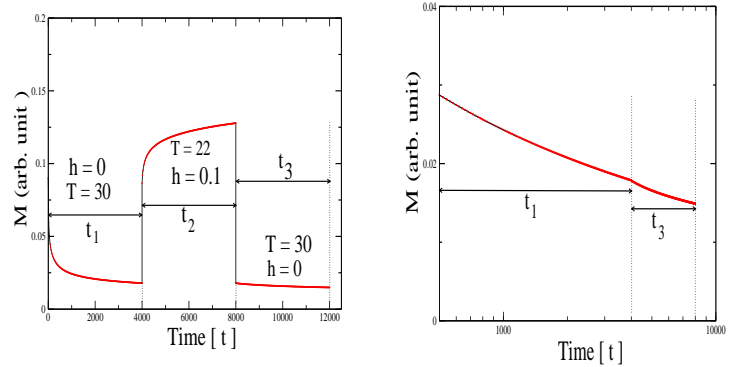}}}}
\caption{(color online) (a) Numerically simulated FC relaxation curves with temporary cooling and field change; (b) the same data vs the total time spent at 30 K on a logarithmic scale.}
\end{figure}
We represent our numerical results for the same protocol in Fig. 13. When the field is cut-off the particles with $T_b\leq30K$ do not contribute to the magnetization . After $t_1$, when the sample is quenched to $22K$ and the field is turned on there is naturally a sudden jump in the magnetization due to the particles with $T_b\leq22K$ which have much higher magnetization than the value just before quenching. As discussed earlier the particles with $T_b>30K$ are not in equilibrium and their relaxation is very slow at $T=22K$, which explains an almost constant curve during $t_2$. After $t_2$, the field is turned off, and the temperature is turned back to $T_0$. Naturally, the magnetization jumps down, because the particles with $T_b\leq30K$ reach a new equilibrium state which has almost zero magnetization immediately following the field and temperature changes and the system returns back to its state prevailing before quenching.\\
\indent
Finally, Sun et al have studied magnetic relaxation after a temporary heating (instead of temporary cooling) from 30K to 38K which do not exhibit any memory effect. After temporary heating, when temperature returns back to $T_0$, the system does not come back to its previous state before heating (Fig. 14). Sun et al suggested that this asymmetric response with respect to negative/positive temperature cycling is consistent with a hierarchical model of the spin-glass phase. However, we have numerically reproduced the same results as that of Sun et al based on our two-state independent particle model, as shown in Fig. 15. No memory effect appears after positive heating  which can be explained as follows.
\begin{figure}[h]
{\rotatebox{0}{\resizebox{8cm}{8cm}{\includegraphics{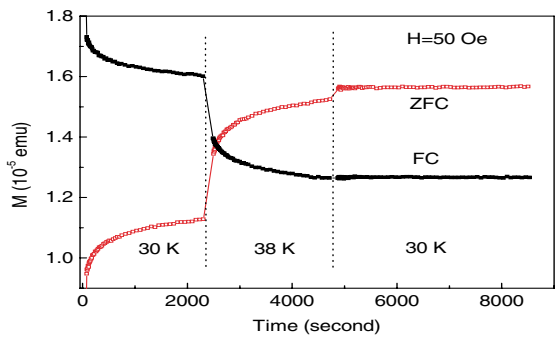}}}}
\vskip-1.5cm
\caption{(color online) Magnetic relaxation measurements in $Ni_{81}Fe_{19}$ with temporary heating for the ZFC and FC methods. (Sun et al Phys. Rev. Lett. {\em 91}, 167206 (2003)).}
\end{figure}
In the FC method the sample is cooled to $T_0=30K$ in the presence of a 50 Oe field and then the field is cut-off and the relaxation is measured for a time $t_1$. So the magnetization decreases with time for a time $t_1$. Now as the temperature is increased all the particles with $T_b\leq38K$ respond to this temperature change and relax to the new equilibrium state. Since thermal agitation increases with the increase of temperature, magnetization decreases further for the time $t_2$. As the temperature returns back to $T = 30 K$, the particles with $T_b>30K$ are unable to respond to this temperature change. Thus the particles with $T_b\leq30K$ actually follow the path during time $t_2$ rather than $t_1$. Because all the particles which had flipped during the time $t_1+t_2$ cannot return back to their previous state as that before heating, no memory effect is observed.\\
\begin{figure}[h]
{\rotatebox{270}{\resizebox{8cm}{8cm}{\includegraphics{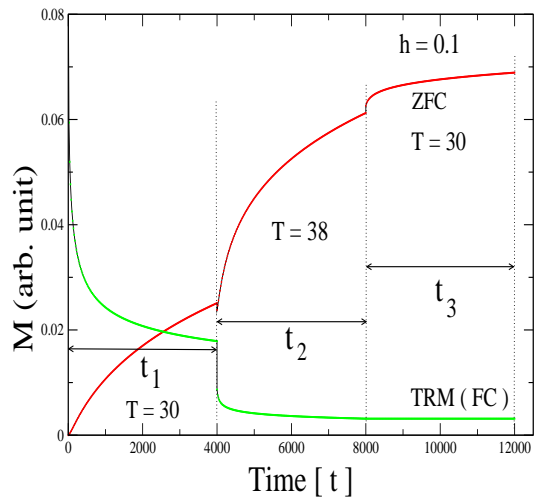}}}}
\caption{(color online) Numerically simulated FC and ZFC relaxation curves with temporary heating.}
\end{figure}
In the ZFC method the sample is cooled to $T_0$ in the absence of an external field and then a 50 Oe field is turned on and relaxation is measured for a time $t_1$, yielding a finite magnetization, for particles with $T_b\leq 30K$. Then a slow logarithmic relaxation begins which is due to the energy distribution of the particles. As the sample is further heated to $T = 38K$, all the particles with $T_b=38K$ respond to this temperature change. Thus the logarithmic relaxation is continued but there is a jump in magnetization, because the particles with $T_b\leq38K$ and those flipped during $t_1$ reach a new equilibrium state. As the temperature of the sample is returned back to $T=30K$ thermal agitation is reduced, so there is a jump in magnetization. But now only the particles with $T_b>30K$ are allowed to relax and their relaxation is very slow at $T=30K$, so we obtain an almost flat curve.\\
\indent
We conclude this section by underscoring that our simulations based on the simple two-state noninteracting model reproduce all the features of the memory effects observed by Sun et al in the Permalloy ($Ni_{81}Fe_{19}$). Secondly, positive heating does not yield memory effect whereas temporary cooling does. So there is an asymmetric response with respect to negative/positive temperature cycling. This asymmetry is due to the fact that after temporary cooling only smaller nanoparticles are able to respond to the temperature or field  change and relax to the new equilibrium state. The larger nanoparticles are frozen. Upon returning to the initial temperature or field value, the smaller particles rapidly respond to the change such that this new state is essentially the same as that before the temporary cooling, and the larger nanoparticles are now able to resume relaxing to the equilibrium state. This results in a continuation of the magnetic relaxation after the temporary temperature or field change. In contrast, for  positive heating, all the particles smaller as well as bigger are able to respond to the temperature or field change. Therefore, after returning to the initial temperature, the bigger particles do not respond at all whereas the smaller particles take time to respond, thus leading to no memory effect in the positive heating cycle.\\ 
{\section {Spinglass Like slow dynamics}}
 Time-dependent magnetization measurements suggest that dense nanoparticle samples may exhibit glassy dynamics due to dipolar inter-particle interaction \cite{gil,luo,mamiya,jonsson}; disorder and frustration are induced by the randomness in the particle positions and anisotropy axis distributions.\\
\vskip-2.0cm
\begin{figure}[h]
{\rotatebox{0}{\resizebox{10cm}{10cm}{\includegraphics{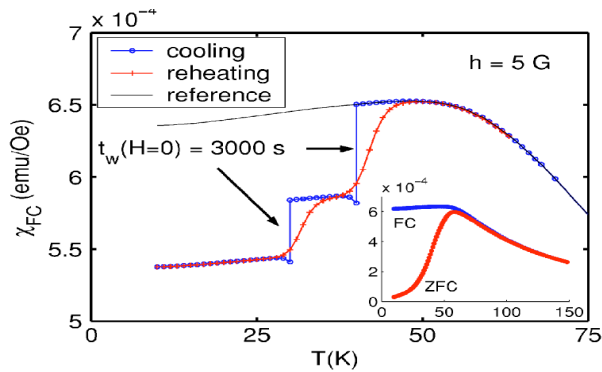}}}}
\vskip-3.0cm
\caption{(color online) FC susceptibilities of the Fe-N system with the same protocol as described in the text for the double memory experiment (DME). Inset shows the FC and ZFC susceptibilities vs. temperature of the Fe-N system. (Sasaki et al Phys. Rev. B {\em 71}, 104405 (2005)).}
\end{figure}
\begin{figure}[h]
\hskip -1.4cm
{\rotatebox{0}{\resizebox{10cm}{6cm}{\includegraphics{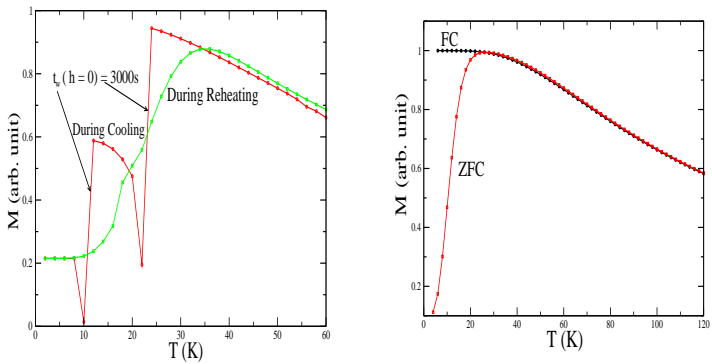}}}}
\caption{(color online) (a) Numerically simulated results for the double memory experiment (DME); (b) the FCM and ZFCM vs. temperature of the interacting system are shown.}
\end{figure}
Figure 16 shows the ZFC and FC behavior (for the linear susceptibility, which is simply proportional to the magnetization) in the dense magnetic nanoparticle system of Fe-N, measured by Sasaki et al \cite{sasaki}. For comparison, our numerical results are shown in Fig. 17. We observe a peak in the ZFCM which corresponds to an average blocking temperature $<T_b>$. In the superparamagnetic case the ZFC-FC curves bifurcate at a temperature away from the peak position of the ZFCM (see Fig. 1). On the other hand, for the dense system the ZFCM-FCM curves bifurcate at a temperature very close to the peak position of the ZFCM. The FCM of the dense system does not increase but stays almost constant below  $<T_b>$ which is the primary indicator for the glassy state \cite{sasaki}. It is interesting that we have been able to numerically reproduce the same kind of FC-ZFC curves based on our simple-minded interacting nanoparticle model, summarized at the end of Sec. II.\\
\begin{figure}[h]
{\rotatebox{0}{\resizebox{8cm}{8cm}{\includegraphics{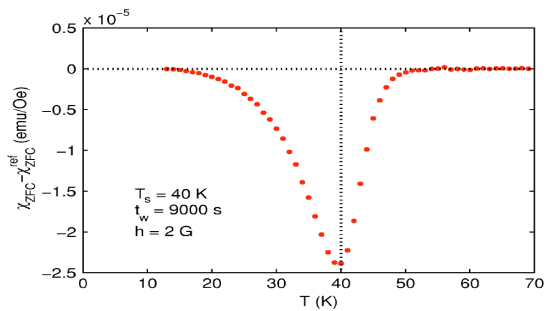}}}}
\vskip-1.0cm
\caption{(color online) Difference of the ZFC susceptibilities of the Fe-N system vs. temperature. (Sasaki et al Phys. Rev. B {\em 71}, 104405 (2005)).}
\end{figure}
\begin{figure}[h]
{\rotatebox{270}{\resizebox{8cm}{8cm}{\includegraphics{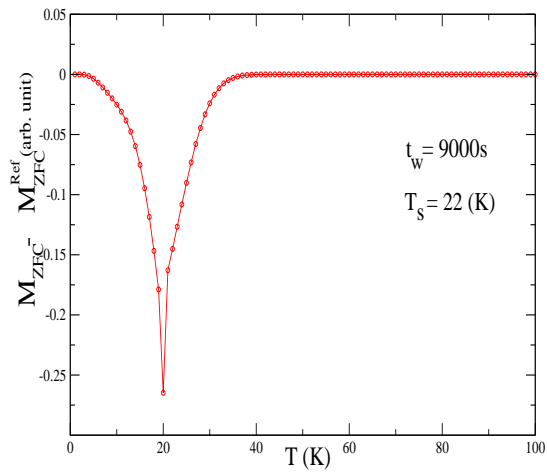}}}}
\caption{(color online) Numerically simulated results for the memory experiment (DME) for the ZFC method in interacting system.}
\end{figure}
In order to have a better understanding of glassy relaxation, time-dependent magnetization studies under various heating and cooling protocols were performed by Sasaki et al \cite{sasaki} on dense Fe-N nanoparticle systems, by Raj Sankar et al \cite{raj} on $LaMnO_{3.13}$, by Kundu et al on $La_{0.7}Ca_{0.3}CoO_{3}$ and by Telem-Shafir and Markovich on MICS76 sample \cite{gil}. Figure 16 also shows the results of double memory experiment (DME) under FC protocol by Sasaki et al.  In this, the system is cooled under a field of 50 Oe. The field is cut-off during the intermittent stops of the cooling at $T=30K$ and at $T=40K$ for 3000 sec at each stop. After reaching the lowest temperature the susceptibility measurement is repeated in the heating mode without any intermittent stop. In Fig. 17 we have shown our numerically simulated results of DME based on our interacting nanoparticle model, which have a striking resemblance to the experimental results.\\
\indent
Another protocol has been suggested by Sasaki et al to confirm whether the observed memory effect is due to glassy behavior or not. In this experiment the sample is first rapidly cooled in zero field from a reference temperature ($T_{ref}$) to the stop temperature ($T_s$), where it is kept for 9000 sec. The cooling is then resumed down to the lowest temperature where a magnetic field is applied and the susceptibility is recorded on reheating the sample. The conventional ZFC susceptibility is also recorded. The difference between the aged and the normal ZFC susceptibility as a function of temperature is shown in Fig. 18. Figure 19 is our numerically simulated results, which are again very similar to that of experimental results. In all the above mentioned simulations we have used a log-normal distribution of particle sizes wherein the parameter $\gamma$ is set to 0.5. In the fitting procedure the average anisotropic energy $K\bar{V}$ is chosen as the unit of energy as well as that of temperature by setting $k_B=1$ and $\bar{V} = \exp(\gamma^2/2)$. Because $\tau_0$ for nanoparticles is around $10^{-9}$ sec and a typical experimental time window is about 10 sec, we have investigated the predictions of our model in the time window $10^{10}\tau_0$. To introduce the dipolar interaction we have numerically solved the self-consistent Eq. (14) for H and inserted this value in Eq. (15) to calculate rate constants and hence the magnetization. The external field is set to 0.1, whereas $\Lambda$ has been taken as a random variable between 0 and 1. In conclusion, we find that a large class of data on memory effects in nanomagnetic systems can be explained by a simple rate theory, without and with interactions, in a meanfield approximation.\\ 
\\
\section*{Acknowledgments}
M.B. acknowledges financial support from the Council of the Scientific and Industrial Research (CSIR), Government of India. SD is grateful to Prof. G. Markovich and Prof. C. N. R. Rao for helpful discussions on their experimental results during an Indo-Israel symposium on nanosciences.

\end{document}